\documentclass[pre,amsmath,amssymb,twocolumn]{revtex4-2}
\usepackage[utf8]{inputenc}
\usepackage[T1]{fontenc}
\usepackage{graphicx}
\usepackage{color}
\usepackage{bm}
\usepackage[hidelinks]{hyperref}

\renewcommand*{\vec}[1]{\mathbf{#1}}
\newcommand*{\diff}{\mathop{}\!\mathrm{d}}

\newcommand{\rv}{{\vec{r}}}

\newcommand{\rhotwo}{\rho_{2}}

\graphicspath{{./figures/}}

\begin{document}

\title{Comparative study of force-based classical density functional theory}

\author{Florian Sammüller}
\author{Sophie Hermann}
\author{Matthias Schmidt}
\email{Matthias.Schmidt@uni-bayreuth.de}
\affiliation{Theoretische Physik II, Physikalisches Institut, Universität Bayreuth, D-95447 Bayreuth, Germany}

\date{\today}

\begin{abstract}
  We reexamine results obtained with the recently proposed density functional theory framework based on forces (force-DFT) [\citeauthor{Tschopp2022}, Phys.\ Rev.\ E \textbf{106}, 014115 (2022)].
  We compare inhomogeneous density profiles for hard sphere fluids to results from both standard density functional theory and from computer simulations.
  Test situations include the equilibrium hard sphere fluid adsorbed against a planar hard wall and the dynamical relaxation of hard spheres in a switched harmonic potential.
  The comparison to grand canonical Monte Carlo simulation profiles shows that equilibrium force-DFT alone does not improve upon results obtained with the standard Rosenfeld functional.
  Similar behavior holds for the relaxation dynamics, where we use our event-driven Brownian dynamics data as benchmark.
  Based on an appropriate linear combination of standard and force-DFT results, we investigate a simple hybrid scheme which rectifies these deficiencies in both the equilibrium and the dynamical case.
  We explicitly demonstrate that although the hybrid method is based on the original Rosenfeld fundamental measure functional, its performance is comparable to that of the more advanced White Bear theory.
 \end{abstract}

\maketitle

\section{Introduction}

Whether any theoretical approach is useful in practice often stems from the accuracy and reliability of its predictions versus the analytical and computational effort it requires.
Classical density functional theory (DFT) \cite{Evans1979,Hansen2013} fares very well, ranging from simple local density and square gradient approximations \cite{Evans1979,Hansen2013}, which are sufficiently accurate in appropriate circumstances (see e.g.\ Refs.\ \onlinecite{Eckert2021,Eckert2022,Eckert2022a} for studies of colloidal sedimentation) to the nonlocal and nonlinear prowess of Rosenfeld's fundamental measure theory (FMT) \cite{Rosenfeld1989,Roth2010} to capture hard sphere correlations.

Applying DFT in practice involves solving a variational (minimization) problem, which typically requires the numerical treatment of an implicit integral equation.
One obtains static quantities or performs adiabatic time evolution within dynamical DFT (DDFT).
The later task is often done with a simple time-forward integrator, but more advanced methods \cite{Aduamoah2022,Roden2022} allow to address dynamical optimization problems.
Similarly, computational grids in real space range from simple and often very relevant effective one-dimensional geometries \cite{Davidchack2016} to full three-dimensional resolution \cite{Stopper2017} and pseudo-spectral methods \cite{Aduamoah2022}.
Increasing the complexity of the underlying microscopic model trades off well with the achieved broader physical scope, as is the case in including orientational degrees of freedom in liquid crystal formation \cite{Heras2004,Heras2005} and molecular DFT \cite{Zhao2011,Jeanmairet2013,Jeanmairet2020} for realistic modeling of molecular liquids.

DFT offers a complete theoretical framework for addressing static problems in many-body statistical physics.
The theory is founded on the concept of potentials, including the chemical potential $\mu$ as a control parameter, an external potential that adds local variation to $\mu$, and an intrinsic part, which arises from the interparticle interactions and which induces the coupling of the microscopic degrees of freedom.

In contrast to this basis in potentials, the concept of forces seems almost alien to the framework, or at least redundant.
Nevertheless, in a variety of very different fields there appears to be new interest in this old workhorse.
We mention the recent and unexpected advances in simulation methodology based on force-sampling \cite{Rotenberg2020,Heras2018,Purohit2019,Borgis2013} and in the related but different realm of quantum DFT \cite{Tokatly2005,Tokatly2005a,Tokatly2007,Tchenkoue2019,Tarantino2021}, as well as in the power functional approach to nonequilibrium many-body dynamics \cite{Schmidt2022}.
Both the classical and the quantal force balance were proven to be direct consequences of a thermal Noether symmetry of the system \cite{Tschopp2022,Hermann2022}.
Forces are also central in the recent treatment of motility-induced phase separation by Brady and coworkers \cite{Omar2022}.

Recently \citeauthor{Tschopp2022} \cite{Tschopp2022} developed a force-based alternative to implement density functional theory.
Their ``force-DFT'' comes at an increased computational cost, as two-body functions appear explicitly and need to be manipulated.
Nevertheless, the framework still retains formal one-body purity with the two-body density playing the role of an auxiliary variable.
The difference between the standard approach to DFT and the force-DFT appears similar to the difference between the virial and compressibility route to determine the equation of state in bulk fluids \cite{Hansen2013}, e.g.\ on basis of the celebrated Percus-Yevick approximation for the hard sphere fluid.
Actually, as could be shown by \citeauthor{Tschopp2022} \cite{Tschopp2022} via an investigation of the hard wall contact theorem, standard DFT corresponds in this case to the compressibility equation of state while force-DFT satisfies the virial equation of state.

Here we address the question of where the balance of complexity and accuracy tips for the force-DFT.
We compare the theoretical results of Ref.\ \onlinecite{Tschopp2022} against new computer simulation data, involving canonical, grand-canonical, and event-driven methods, as is appropriate for carrying out a systematic comparison, as we detail below.
We find that the force-DFT per se does not improve on standard DFT in the considered cases, but that an appropriate linear combination of results from the two approaches, which constitutes a simple hybrid scheme, gives much improved results as compared to the standard framework.
We hence follow the suggestion raised in the outlook of Ref.\ \onlinecite{Tschopp2022} that ``the virial and compressibility routes could be mixed in the spirit of liquid-state integral-equation theories, using approximations analogous to the Rogers-Young or Carnahan-Starling theories.''

The paper is organized as follows.
In Sec.\ \ref{sec:theory_overview}, a brief summary of the core concepts of DFT is given.
Particularly, we highlight the conceptual differences of the force-DFT approach and describe how both routes can be used to formulate a dynamical DFT.
In Sec.\ \ref{sec:comparison}, we conduct a thorough reinvestigation of the force-DFT results for the model applications of Ref.\ \onlinecite{Tschopp2022}, thereby comparing this data to results from standard DFT and from simulation.
Throughout this work, the hard sphere fluid is considered and the force-DFT results are those that were obtained with the Rosenfeld \cite{Rosenfeld1989} FMT functional in Ref.\ \onlinecite{Tschopp2022}.
We first turn to the case of imposing a planar hard wall in Sec.\ \ref{sec:comparison_equilibrium} where the respective connection of standard and force-DFT to the compressibility and virial route is established via the hard wall contact theorem.
To obtain numerically accurate results for this equilibrium situation, we perform grand canonical Monte Carlo (GCMC) \cite{Frenkel2001} simulations which are systematically adjusted to enable a comparison with both DFT routes.
In Sec.\ \ref{sec:comparison_dynamics}, the dynamical behavior of the hard sphere fluid in a switched harmonic potential is considered.
For the numerical reproduction of the exact time evolution, we employ event-driven Brownian dynamics simulations (EDBD) \cite{Scala2007} that are initialized with particle configurations from canonical Monte Carlo (MC) simulation.
The time-dependent density profile obtained with this procedure is compared to results from standard and from force-DDFT.
Based on the observations of Sec.\ \ref{sec:comparison_equilibrium} and \ref{sec:comparison_dynamics}, we investigate a hybrid scheme in Sec.\ \ref{sec:hybrid} as a means to substantially improve the resulting density profiles via a linear combination of results from the standard and force route.
This is illustrated both for the equilibrium and for the dynamical case, where we find much better agreement with simulation results.
In particular, we show that hybrid Rosenfeld DFT can compete with standard DFT on the basis of the high-accuracy White Bear \cite{Roth2002,HansenGoos2006} functionals for the hard wall test case.
We conclude in Sec.\ \ref{sec:conclusion} and give an outlook to further possible applications of force-DFT and the hybrid scheme.

\section{Concepts of standard DFT and force-DFT}
\label{sec:theory_overview}

One of the main goals and motivations behind the development of force-DFT is the possibility to improve upon the results from standard DFT calculations.
Usually improvements of DFT involve refinements of the assumed free energy density functional.
Two prominent examples are the advanced White Bear versions of FMT \cite{Roth2002,HansenGoos2006,Roth2010}.
In contrast, the implementation of force-DFT acknowledges the fact that the exact density functional is not within reach for relevant physical systems and that intoducing approximations leads to a theory that is not entirely self-consistent.
Starting from the same functional but using different routes to calculate a physical variable will yield different results except in the formal case of an exactly known functional.

The starting point of both the standard DFT and the force-DFT approach is determining the density $\rho(\rv)$ self-consistently from solving the Euler-Lagrange equation
\begin{align}
\ln\rho(\rv) - \beta (\mu-V_\text{ext}(\rv)) -  c_{1}(\rv) = 0, \label{eq:el}
\end{align}
where $\beta=(k_\text{B}T)^{-1}$ denotes the inverse temperature with $k_B$ being Boltzmann's constant, and $\mu$ is the chemical potential.
While the thermodynamic state point as well as the external potential $V_\text{ext}(\rv)$ act as control parameters, the one-body direct correlation function $c_1(\rv)$ arises from internal interactions and it has to be approximated in practice.

Given a suitable approximation for the excess free energy density functional $F_\text{exc}[\rho]$, where the brackets indicate functional dependence, one determines the one-body direct correlation function via functional differentiation according to
\begin{align}
c_{1}(\rv) = - \beta \frac{\delta F_\text{exc}[\rho]}{\delta \rho(\rv)}. \label{eq:c1p}
\end{align}
In force-DFT one retains eq.\ \eqref{eq:el} but calculates the direct correlation function from the force integral
\begin{align}
c_{1}(\rv) =  -\nabla^{-1} \cdot \int d \rv' \frac{\rhotwo(\rv, \rv';[\rho])}{\rho(\rv)} \nabla \beta\phi(|\rv- \rv'|), \label{eq:c1f}
\end{align}
where $\nabla^{-1} \!=\! 1/(4\pi) \int d \rv' (\rv-\rv') / |\rv-\rv'|^3$ indicates an integral operator (see e.g.\ Refs.\ \onlinecite{Heras2018,Rotenberg2020}) and $\phi(r)$ is the pair interaction potential as a function of the interparticle distance $r$.
At face value the expression \eqref{eq:c1f} is based on the two-body level as it depends on the two-body density $\rhotwo(\rv,\rv';[\rho])$.
However, starting from an approximative excess free energy functional $F_\text{exc}[\rho]$, the two-body density $\rhotwo(\rv,\rv';[\rho])$ is determined by functionally differentiating twice to get the two-body direct correlation function $c_2(\rv,\rv')=-\beta \delta F_\text{exc}[\rho]/\delta \rho(\rv)\delta \rho(\rv')$ and then solving the inhomogeneous Ornstein-Zernike (OZ) equation self-consistently \cite{Tschopp2022}.
The last step can be done numerically in planar and spherical geometry, see Refs.\ \onlinecite{Goetzelmann1996,Tschopp2020,Tschopp2021} for the technical details.

Solving the inhomogeneous OZ equations has relevant applications in the study of the structure factor of thin films \cite{Klimpel1999}, of capillary waves and of the wave-number dependent surface tension \cite{Mecke1999,Hoefling2015} in lateral systems.
Due to this additional self-consistency step and by working on the two-body level, the force-DFT is technically and computationally more complex than standard implementations of DFT based on eq.\ \eqref{eq:c1p}.

The alternative force route also transfers directly to DDFT, which is then called force-DDFT.
Standard DDFT provides a statistical mechanical approach to describe inhomogeneous fluids in nonequilibrium, including the dynamics of adsorption \cite{AngiolettiUberti2014,AngiolettiUberti2018}, lane formation \cite{Chakrabarti2003,Chakrabarti2004} or the motion of active microswimmers \cite{Menzel2016,Sharma2017} (see the review \cite{Vrugt2020} for a recent and broad overview).
This theory is the dynamic extension of DFT and it is intrinsically based on the adiabatic approximation.
Efforts to improve the implied approaches \cite{Vrugt2022} include the in principle exact power functional theory, which goes beyond the adiabatic approximation by taking all superadiabatic (above adiabatic) contributions into account \cite{Schmidt2013,Schmidt2022}.
Recently, a concrete implementation of a two-body DDFT \cite{Tschopp2022a}, which is deeply founded on the force route investigated in this work, has been shown to incorporate superadiabatic effects on the one-body level, thus providing a way improve upon standard DDFT.
Ref.\ \onlinecite{Heras2023} discusses the shortcomings of standard DDFT and describes possible ways forward.

The transition from the equilibrium DFT to the nonequilibrium DDFT is in both cases simply based on the continuity equation
\begin{align}
\frac{\partial \rho(\rv,t)}{\partial t} = - \nabla \cdot \vec{J}(\rv,t). \label{eq:continuity}
\end{align}
The current $\vec{J}(\rv,t)$ is equal (up to the friction constant) to the force density and takes into account its internal, external and diffusive ideal gas contribution.
The internal force $\vec{f}_\mathrm{int}(\rv,t)$ is then assumed, as in equilibrium, to be obtained by the gradient of the one-body direct correlation function, $\vec{f}_\mathrm{int}(\rv,t) = k_B T \nabla c_1(\rv,t)$, which neglects superadiabatic force contributions \cite{Schmidt2022}.
Evaluation of $c_1(\rv,t)$ can proceed via eq.\ \eqref{eq:c1p} for the DDFT route and via eq.\ \eqref{eq:c1f} in case of the force-DDFT approach, and differences are expected to occur for approximate forms of the excess free energy functional.

\section{Comparison to simulation results}
\label{sec:comparison}

\subsection{Equilibrium: hard sphere fluid at a hard wall}
\label{sec:comparison_equilibrium}

\begin{figure}[htb]
  \centering
  \includegraphics{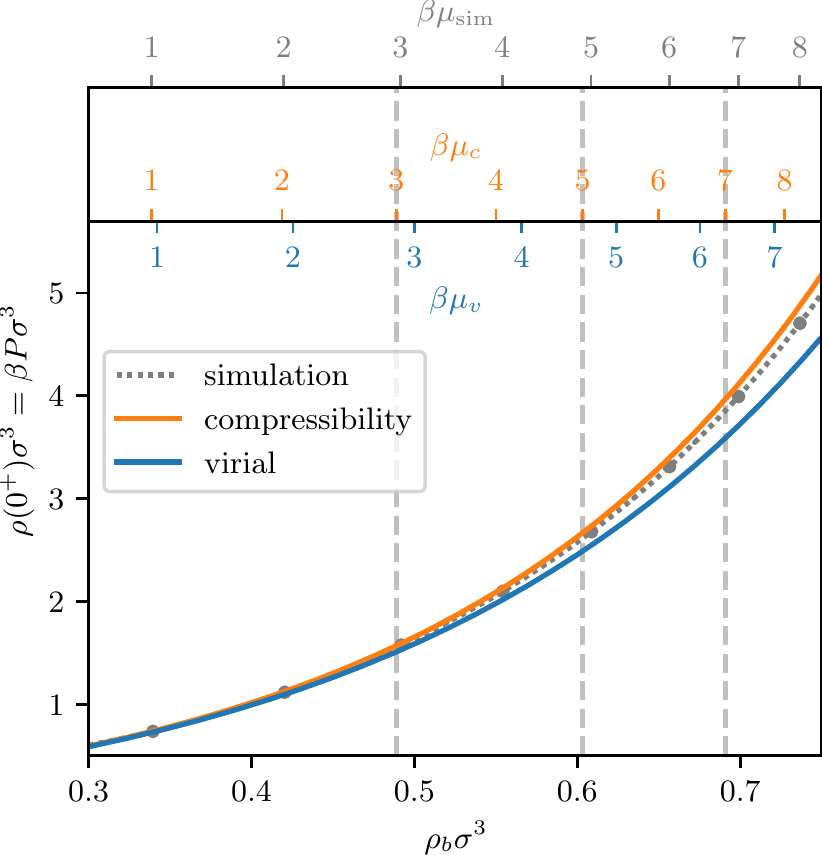}
  \caption{
    The equation of state of the hard sphere fluid is shown as obtained from the Percus-Yevick approximation both via the compressibility and the virial route as well as from GCMC simulations (as indicated).
    Thereby, $\rho_b$ denotes the bulk density and $\rho(0^+) = \beta P$ is the contact density at the hard wall, which can be associated with the bulk pressure $P$.
    The upper scales illustrate differences in the chemical potential with respect to the simulation values $\mu_\mathrm{sim}$ that result from the approximative equations of state via the compressibility ($\mu_c$) and virial ($\mu_v$) route (analytical expressions are given in Appendix \ref{appendix:mu}).
    Therefore, to yield a valid comparison of the density profiles, $\mu$ has to be tuned appropriately in the GCMC simulation to match the considered bulk densities of the standard and force-DFT results, which is illustrated by the gray vertical lines.
  }
  \label{fig:eos}
\end{figure}

\begin{figure}[!htb]
  \centering
  \includegraphics{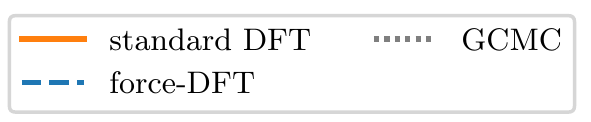}\\
  \vspace{0.1in}
  \includegraphics{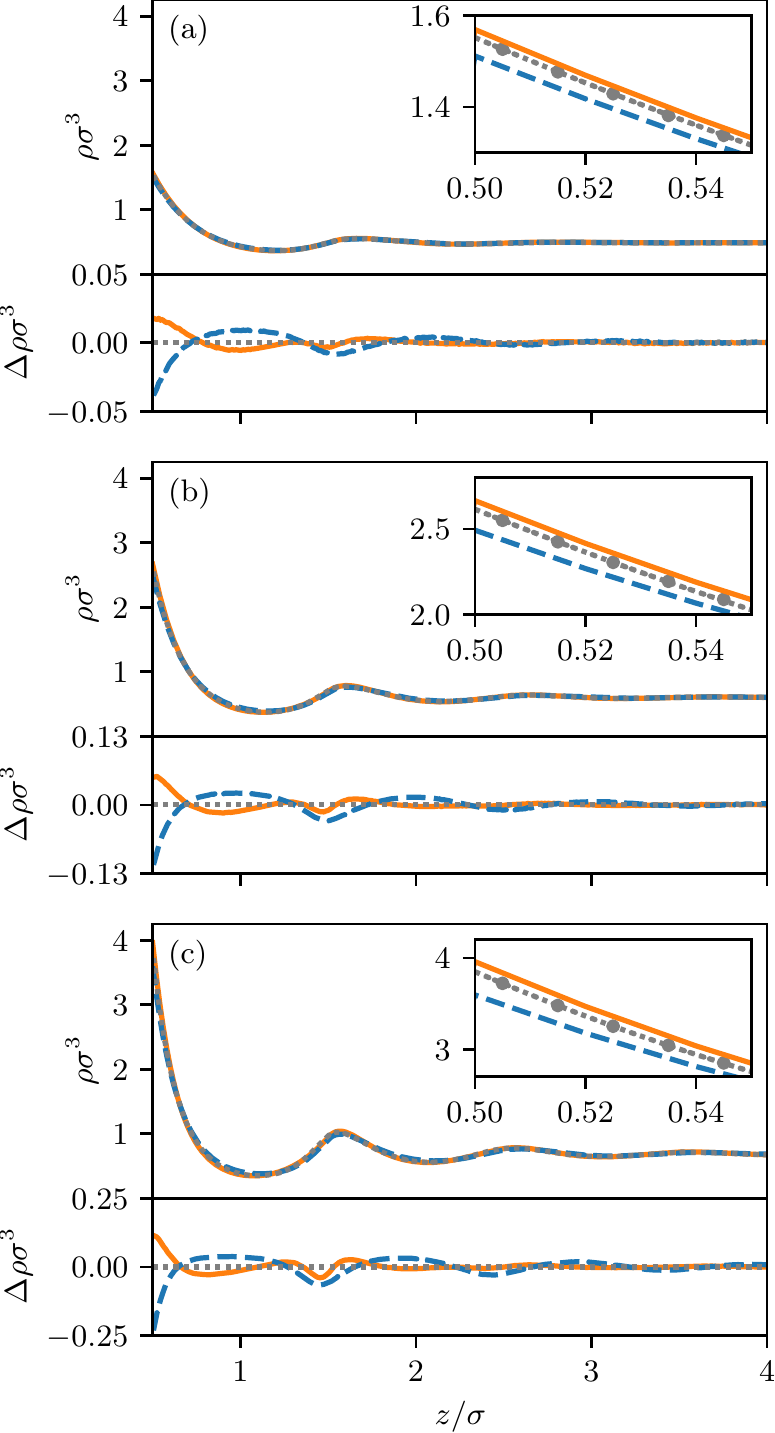}
  \caption{
    Density profiles $\rho(z)$ of a hard sphere fluid at a planar hard wall are shown for values $\rho_b = 0.4890 \sigma^{-3}$ (a), $\rho_b = 0.6032 \sigma^{-3}$ (b) and $\rho_b = 0.6908 \sigma^{-3}$ (c) of the bulk density.
    We compare the results of standard (orange) and force-DFT (blue) to numerically exact density profiles from GCMC simulations (gray).
    For each value of $\mu$, the absolute error $\Delta \rho(z)$ of the density profiles compared to the simulation result is shown in the respective bottom panel, and the inset plot zooms in on the differences of the two DFT routes close to the hard wall.
    The simulations were set up to yield the same bulk density as in the DFT results via an appropriate choice of the chemical potential (cf.\ fig.\ \ref{fig:eos} and Table \ref{tab:hard_wall_mus}) for a systematic comparison of the resulting contact densities.
  }
  \label{fig:hard_wall}
\end{figure}

We proceed with a comparison of results from both DFT routes to simulation data for the standard case of an equilibrium hard sphere fluid at a hard wall as previously investigated by \citeauthor{Tschopp2022} \cite{Tschopp2022}.
For the DFT treatment of the hard sphere fluid, these authors resorted to the Rosenfeld \cite{Rosenfeld1989} fundamental measure theory (FMT) functional for modeling $F_\mathrm{exc}[\rho]$ in both standard and force-DFT.
As this functional is an approximation, we showcase in the following the deviation to numerically exact grand canonical Monte Carlo \cite{Frenkel2001} (GCMC) data.

Imposing a planar hard wall is a conceptually important test case for two reasons.
First, large density inhomogeneities are induced in the vicinity of the wall, which reveal deviations of approximative theories very clearly \cite{Davidchack2016}.
Second, for arbitrary fluids at a hard wall, the contact theorem
\begin{equation}
  \label{eq:contact_theorem}
  \rho(0^+) = \beta P
\end{equation}
establishes a connection of the bulk pressure $P$ of the fluid to the contact value $\rho(0^+)$ of the density profile.
This holds beyond simple fluids as governed by a pair potential, because DFT is formally valid for many-body interparticle interactions.
As was shown in Ref.\ \onlinecite{Tschopp2022}, standard and force-DFT can be associated respectively in this regard to the compressibility and virial route of liquid integral equation theory \cite{Hansen2013}.
More precisely, it could be proven \cite{Tschopp2022} that
\begin{align}
  \label{eq:contact_theorem_c}
  \rho_s(0^+) &= \beta P_c,\\
  \label{eq:contact_theorem_v}
  \rho_f(0^+) &= \beta P_v,
\end{align}
where $\rho_s(z)$ indicates the density profile as obtained from standard DFT, whereas $\rho_f(z)$ is the density profile obtained with force-DFT as a function of the distance $z$ from the wall.
Eqs.\ \eqref{eq:contact_theorem_c} and \eqref{eq:contact_theorem_v} can be derived by explicit analytical calculation and they connect the respective contact densities ($z = 0^+$) to the compressibility ($P_c$) and virial ($P_v$) forms of the pressure which are well-known bulk results from liquid integral equation theory.
The two DFT routes thus make these differences accessible locally and away from the wall on the level of the inhomogeneous density profile.
As the force-DFT is inherently tailored to simple fluids that are governed by pairwise interparticle interactions [recall eq.\ \eqref{eq:c1f}], the force-DFT contact theorem \eqref{eq:contact_theorem_v} also only holds for simple fluids, whereas eq.\ \eqref{eq:contact_theorem_c} is general.
For details of the respective proofs we refer the reader to Ref.\ \onlinecite{Tschopp2022}.

In the present case, the Rosenfeld FMT functional reproduces by construction the Percus-Yevick bulk fluid results.
In particular, we recall \cite{Hansen2013} the compressibility equation of state
\begin{equation}
  \label{eq:eos_py_c}
  P_c = \frac{\rho_b}{\beta} \frac{1 + \eta + \eta^2}{(1 - \eta)^3}
\end{equation}
and the virial equation of state
\begin{equation}
  \label{eq:eos_py_v}
  P_v = \frac{\rho_b}{\beta} \frac{1 + 2\eta + 3\eta^2}{(1 - \eta)^2},
\end{equation}
where $\rho_b$ is the bulk density and $\eta = \rho_b \sigma^3 \pi / 6$ is the packing fraction.
The standard Rosenfeld FMT when evaluated at a constant density gives a free energy which is consistent with $P_c$ \cite{Roth2010}.

In Ref.\ \onlinecite{Tschopp2022}, the comparison was carried out as follows.
First, standard DFT calculations were performed for various values of the reduced chemical potential $\beta \mu = 3, 5, 7$, which respectively corresponds to bulk densities of $\rho_b \sigma^3 = 0.4890, 0.6032, 0.6908$, cf.\ Table \ref{tab:hard_wall_mus}.
Then, corresponding force-DFT calculations were carried out, which were set up to yield identical bulk densities for providing a valid comparison via eqs.\ \eqref{eq:contact_theorem_c} - \eqref{eq:eos_py_v}.
As the control parameter of force-DFT is the mean number of particles $\langle N \rangle$, instead of the chemical potential $\mu$ as is the case in standard DFT, the results for $\langle N \rangle$ obtained from the standard DFT calculations were taken as input for the force-DFT.
With this protocol, it could be verified that the contact densities of standard and force-DFT indeed correspond to the compressibility and virial pressures \eqref{eq:eos_py_c} and \eqref{eq:eos_py_v}, respectively.

For the following investigations via GCMC simulations, we also want to ensure that the bulk densities match the ones chosen in the DFT calculations.
However, as the Percus-Yevick result \eqref{eq:eos_py_c} deviates slightly from the true equation of state, one cannot merely consider a GCMC simulation with the same value of the chemical potential $\mu$ as in the standard DFT case.
Instead, the value of $\mu$ has to be adjusted to obtain the same bulk density as in both DFT routes.
For this, we perform preliminary simulation runs of the system which yield the numerically accurate equation of state for the hard sphere fluid; results are shown in fig.\ \ref{fig:eos}.
This numerical equation of state is interpolated at the desired values for the bulk density, which then yields the target values of chemical potential for the actual comparison runs (the numerical values are given in Table \ref{tab:hard_wall_mus}).

The density profiles from the thus prepared GCMC simulations and their comparison to both standard and force-DFT results are shown in fig.\ \ref{fig:hard_wall}.
It is observed that the deviation of the contact values at the hard wall indeed reflects the inaccuracies of the Percus-Yevick equation of state.
As expected from the bulk results shown in fig.\ \ref{fig:eos}, the GCMC density profile in the vicinity of the wall is enclosed from above and from below by the two DFT profiles.
The standard DFT result thereby agrees better with the simulation data.
At intermediate separations from the wall, both routes are able to capture the inhomogeneities of the density profile with quite reasonable precision.
Although the simulated density profile lies within the two DFT profiles in most parts of the system, there are also regions where the DFT results do not act as a respective upper and lower bound of the true local density.
This is most clearly visible for large values of $\mu$ (e.g.\ in panel (c)) and close to the first density maximum, where both DFT routes underestimate the values of $\rho(z)$ locally.
The shape of the first density maximum of a hard sphere fluid at a hard wall is particularly difficult to reproduce in DFT even when using more elaborate free energy functionals \cite{Davidchack2016,Yu2002,Roth2002,HansenGoos2006} (we return to this point below).
While providing a means to yield an additional approximation of $\rho(z)$, force-DFT is not capable to systematically rectify this deficiency in the considered case of the hard sphere fluid adsorbed against a planar hard wall.

\begin{table}[htb]
  \caption{
    The values of the chemical potential $\mu_\mathrm{sim}$ for the GCMC simulations that yield matching bulk densities $\rho_b$ with the DFT results (cf.\ fig.\ \ref{fig:eos}).
    The reference chemical potentials $\mu_c$ that were used in the standard DFT calculations (corresponding to the compressibility route) are listed as well.
  }
  \label{tab:hard_wall_mus}
  \begin{tabular}{|c|ccc|}
    \hline
    $\rho_b \sigma^3$ & 0.4890 & 0.6032 & 0.6908 \\
    \hline
    $\beta \mu_c$ & 3 & 5 & 7 \\
    $\beta \mu_\mathrm{sim}$ & 2.9572 & 4.8930 & 6.7983 \\
    \hline
  \end{tabular}
\end{table}

\subsection{Dynamics: hard sphere fluid in a switched harmonic trap}
\label{sec:comparison_dynamics}

\begin{figure}[htb]
  \centering
  \includegraphics{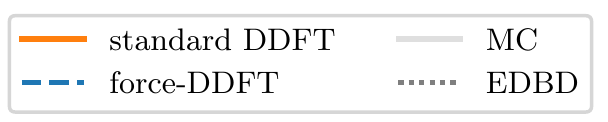}\\
  \vspace{0.1in}
  \includegraphics{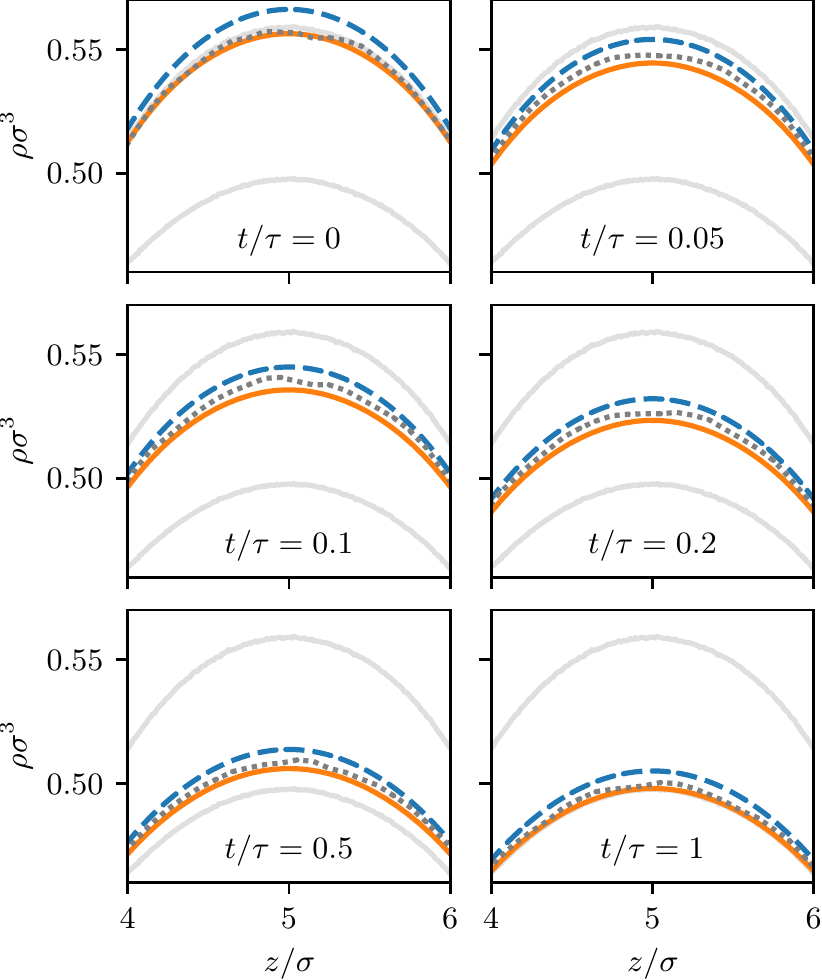}
  \caption{
    Time-evolution of the density profile $\rho(z)$ of a hard sphere fluid in a harmonic external potential $V_\mathrm{ext}(z) = A (z - 5 \sigma)^2$ after switching its strength from $A = 0.75 k_B T / \sigma^2$ to $A = 0.5 k_B T / \sigma^2$ at time $t = 0$.
    The relaxation dynamics calculated with standard (orange) and force-DDFT (blue) are shown for $t / \tau = 0, 0.05, 0.1, 0.2, 0.5, 1$ and are compared to EDBD simulation results (gray).
    The initial and final equilibrium profiles (silver) as obtained via MC simulations for both values of $A$ are indicated in each panel for reference.
  }
  \label{fig:harmonic_trap}
\end{figure}

\citeauthor{Tschopp2022} \cite{Tschopp2022} extended their force-DFT method to out-of-equilibrium situations by replacing the standard form of the one-body direct correlation function $c_1(\vec{r})$ by the force integral \eqref{eq:c1f} in the DDFT equation of motion.
This yields a dynamical description that is still purely adiabatic, i.e.\ it approximates the time-evolution of the system as a series of equilibrium states.
Nevertheless, due to the discrepancies of the two forms of $c_1(\vec{r})$ for a given approximate Helmholtz free energy functional, the two routes will in general lead to different dynamical behavior.
This has been exemplified in Ref.\ \onlinecite{Tschopp2022} for the model situation of a hard sphere fluid in a harmonic external potential $V_\mathrm{ext}(z) = A (z - 5 \sigma)^2$, where the strength of the harmonic trap is switched from $A = 0.75 k_B T / \sigma^2$ to $A = 0.5 k_B T / \sigma^2$ at the initial time $t = 0$.

For a precise numerical investigation of the true time-evolution of the system, we employ event-driven Brownian dynamics (EDBD) simulations \cite{Scala2007}.
Unlike in the equilibrium hard wall comparisons, where the bulk densities of the simulations and the DFT routes were matched to focus solely on structural differences, we now set the total number of particles per lateral system area equal to the corresponding values of the DDFT calculations.
Therefore, differences that arise solely from inaccuracies of the associated equations of state are expected and will be most prominent at the center of the trap, where the local density is large.
To achieve an accurate and fast initialization of each EDBD run, a preliminary canonical Monte Carlo simulation with identical system parameters is carried out, by which appropriately distributed particle configurations of the initial equilibrium state are obtained.
In total, $10^4$ EDBD runs are initialized with the above configurations, and the relaxation dynamics after the switching of the harmonic trap is simulated for $0 \leq t/\tau \leq 1$ with the Brownian timescale $\tau = \sigma^2 \gamma / k_B T$ where $\gamma$ is the friction coefficient.
The time-evolution of the density profile, attained as an average over all runs, is shown in fig.\ \ref{fig:harmonic_trap} for $t / \tau = 0, 0.05, 0.1, 0.2, 0.5, 1$.
Additionally, density profiles for the initial and for the final equilibrium states as obtained via canonical Monte Carlo simulations are depicted.

It is apparent that discrepancies which stem from the approximative form of $F_\mathrm{exc}[\rho]$ emerge for the two DDFT routes.
In the considered system, force-DDFT generally yields larger densities at the center of the harmonic trap.
For the initial and final equilibrium state, standard DFT provides more accurate results in this region.
After toggling the strength of the harmonic potential, both DDFT methods yield similar relaxation dynamics towards their respective equilibrium state.
Compared to the simulation results, the density relaxation is marginally too fast in both routes, as is visible especially shortly after switching the potential (cf.\ fig.\ \ref{fig:harmonic_trap}, $t / \tau = 0.05, 0.1, 0.2$).
This is indicative of nonequilibrium forces that go beyond the adiabatic approximation \cite{Treffenstaedt2020,Schmidt2022} and that are neither captured in standard nor in force-DDFT.

\section{Hybrid scheme}
\label{sec:hybrid}

\begin{figure}[!htb]
  \centering
  \includegraphics{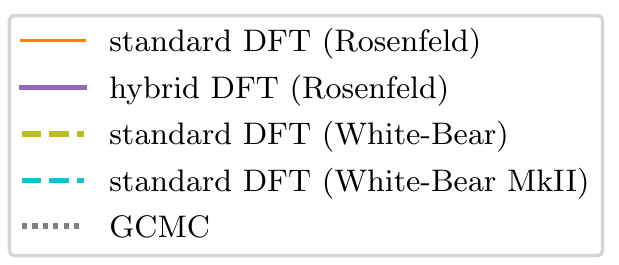}\\
  \vspace{0.1in}
  \includegraphics{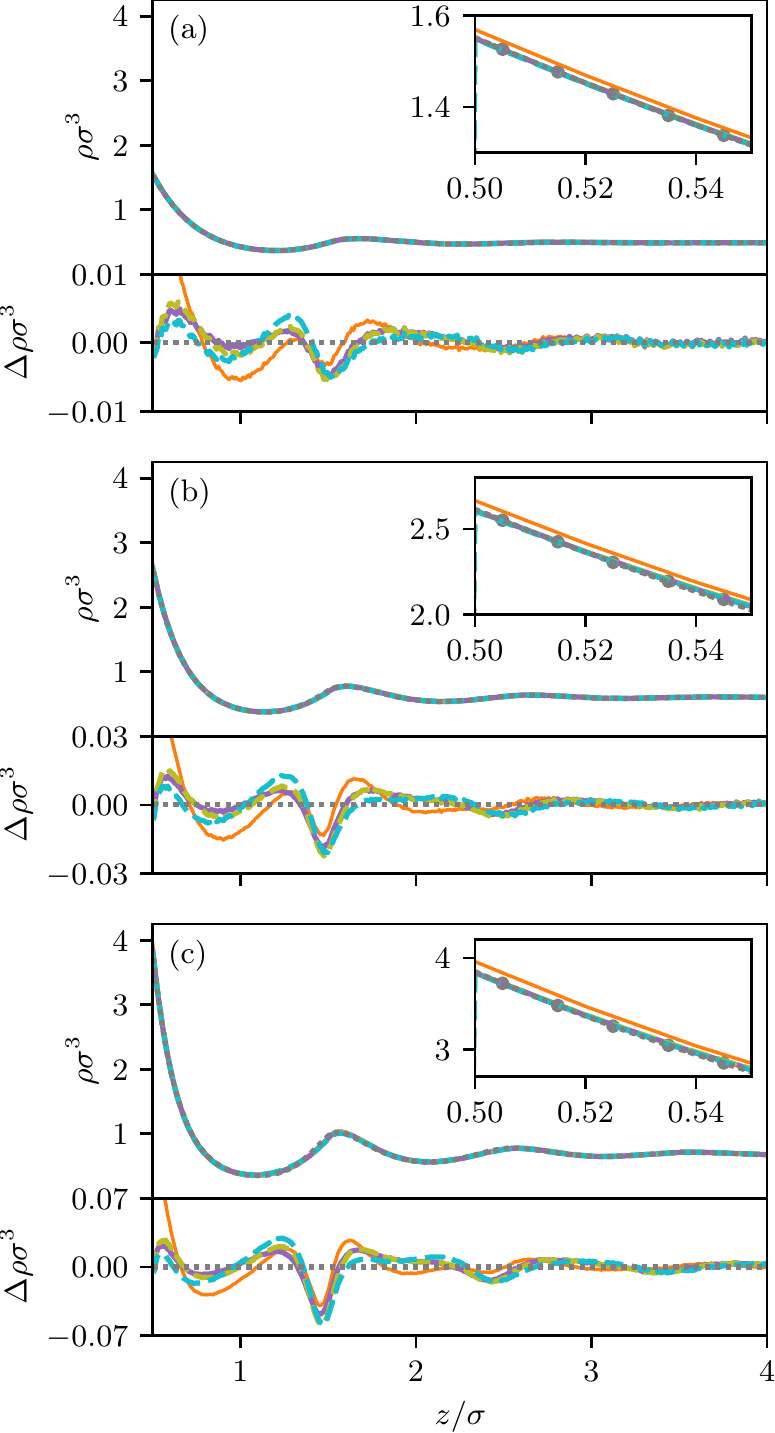}
  \caption{
    Hybrid DFT density profiles $\rho(z)$ (purple) for a hard sphere fluid at a hard wall are compared to simulation results as in fig.\ \ref{fig:hard_wall} (the standard Rosenfeld DFT is replotted in orange).
    In most parts of the system, this combination of standard and force-DFT via eq.\ \eqref{eq:rho_hybrid} enables a systematic improvement of the resulting density profile while retaining the Rosenfeld FMT treatment of $F_\mathrm{exc}[\rho]$.
    The largest discrepancy to the numerical GCMC density profiles (gray) still occurs in the vicinity of the first density maximum.
    For comparison, standard DFT results for the superior White Bear (olive) and White Bear MkII (cyan) functionals are depicted, and an error comparable to hybrid Rosenfeld DFT is found.
  }
  \label{fig:hard_wall_hybrid}
\end{figure}

\begin{figure}[htb]
  \centering
  \includegraphics{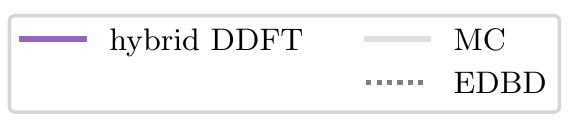}\\
  \vspace{0.1in}
  \includegraphics{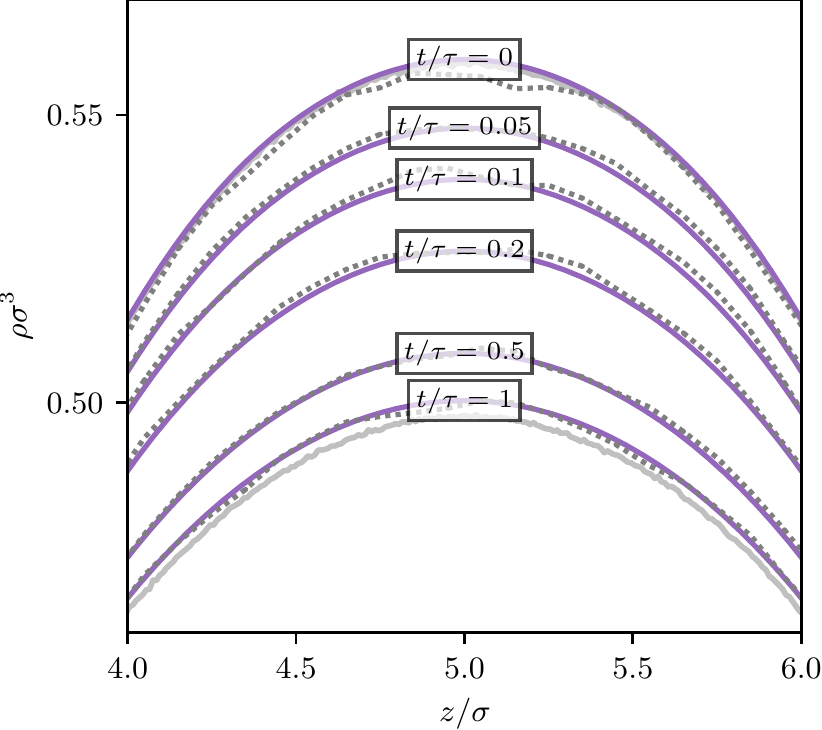}
  \caption{
    Hybrid DDFT density profiles $\rho(z)$ (purple) via eq.\ \eqref{eq:rho_hybrid} for the relaxation of a hard sphere fluid in a harmonic potential as in fig.\ \ref{fig:harmonic_trap}.
    The time evolution is again compared to EDBD simulation results (gray) and the initial and final equilibrium profiles are indicated for reference (silver).
    As in fig.\ \ref{fig:hard_wall_hybrid}, the combination procedure \eqref{eq:rho_hybrid} of standard and force-DDFT yields much better results than the individual routes alone.
  }
  \label{fig:harmonic_trap_hybrid}
\end{figure}

The above comparison of the force-DFT route to standard DFT and simulations reveals that there is no systematic improvement in the resulting density profiles neither in equilibrium (DFT) nor for the dynamical problem (DDFT) considered.
Instead, force-DFT and force-DDFT can be viewed as an alternative to the standard formalism for calculating the density profile from a given Helmholtz free energy functional.
If this functional is not exact, as is the case for the Rosenfeld FMT functional for the hard sphere fluid, the results of both routes will in general differ, as we have exemplified above.
The comparison also uncovers that the numerically exact simulation results are commonly bracketed by standard and force-results for the considered hard sphere fluids.

In this spirit, a systematic improvement of the density profile both in equilibrium and in the dynamical scenario is conceivable by an appropriate combination of the two routes, which constitutes a hybrid implementation of DFT.
For this, we construct a new approximation of the density profile according to
\begin{equation}
  \label{eq:rho_hybrid}
  \rho_h \equiv \alpha \rho_s + (1 - \alpha) \rho_f,
\end{equation}
where the subscripts indicate the results from the hybrid scheme ($h$), from the standard DFT ($s$) and from the force-DFT ($f$).
The interpolation parameter $\alpha$ can be tuned to favor standard ($\alpha = 1$) or force-DFT ($\alpha = 0$).

To arrive at an appropriate choice of $\alpha$ for the considered hard sphere fluids, we recall the Carnahan-Starling \cite{Carnahan1969} equation of state
\begin{equation}
  \label{eq:eos_cs}
  P^\mathrm{CS} = \frac{\rho_b}{\beta} \frac{1 + \eta + \eta^2 - \eta^3}{(1 - \eta)^3}
\end{equation}
as a superior alternative to the Percus-Yevick results \eqref{eq:eos_py_c} and \eqref{eq:eos_py_v}.
In particular, similar to the combination in eq.\ \eqref{eq:rho_hybrid}, eq.\ \eqref{eq:eos_cs} can be obtained from the compressibility ($P^\mathrm{PY}_c$) and virial ($P^\mathrm{PY}_v$) Percus-Yevick equations of state via the linear combination \cite{Hansen2013}
\begin{equation}
  \label{eq:eos_cs_from_py_combination}
  P^\mathrm{CS} = \frac{2}{3} P^\mathrm{PY}_c + \frac{1}{3} P^\mathrm{PY}_v.
\end{equation}

Due to eq.\ \eqref{eq:eos_cs_from_py_combination} and the connection of standard and force-DFT to the compressibility and virial pressure (cf.\ eqs.\ \eqref{eq:contact_theorem_c} and \eqref{eq:contact_theorem_v}), we choose $\alpha = 2/3$ in the following considerations as a means to obtain improved estimates $\rho_h(\vec{r})$ of the density profile via eq.\ \eqref{eq:rho_hybrid}.

The result of this combination of both DFT methods is shown for the hard sphere fluid in equilibrium at the hard wall in fig.\ \ref{fig:hard_wall_hybrid}.
Note that we do not alter the utilized functional, as the hybrid density profile is obtained consistently from a combination of standard and force-results (cf.\ fig.\ \ref{fig:hard_wall}), which were both acquired with the Rosenfeld functional.
The local error of the hybrid Rosenfeld density profile decreases in large parts of the system and particularly in the vicinity of the hard wall as compared to the error of the density profiles obtained via the individual routes.
Hence, hybrid DFT can be considered as a viable means to improve resulting density profiles while avoiding the often difficult task of refining the Helmholtz excess free energy functional.
We further exemplify this in fig.\ \ref{fig:hard_wall_hybrid} by depicting additionally the density profiles obtained from standard DFT when using the more advanced White Bear \cite{Roth2002} and White Bear MkII \cite{HansenGoos2006} functionals, which serves as a benchmark to a common (and the current de facto standard) DFT treatment of the hard sphere fluid.
Notably, the hybrid scheme yields similar accuracy as compared to these results, albeit being obtained with the inferior Rosenfeld approximation for $F_\mathrm{exc}[\rho]$.
In the vicinity of the first density maximum, the hybrid route is still not capable of mitigating the well-known shortcomings of standard FMT completely.
Surprisingly, however, the density profile calculated via eq.\ \eqref{eq:rho_hybrid} match the numerical density profile equally well as both the White Bear and the White Bear MkII functionals employed in standard DFT, in particular for small distances to the hard wall.
Close to the first maximum, the agreement to simulation is even better for the former than in the standard White Bear and White Bear MkII treatment.
This shows that an appropriate combination of standard and force-DFT via eq.\ \eqref{eq:rho_hybrid} to yield a hybrid method is a viable means to improve deficiencies of an approximate excess free energy functional, and that its impact on the density profile may be as significant as when using a superior functional.
A tangible choice of the interpolation parameter in eq.\ \eqref{eq:rho_hybrid} may be obtained via known results for bulk fluids, e.g.\ by comparison of associated equations of state.
While this choice was made analytically with eq.\ \eqref{eq:eos_cs_from_py_combination} for the hard sphere fluid above, bulk simulation results might provide guidance to go beyond Carnahan-Starling results or to apply the hybrid scheme to other particle models.

For the dynamical case, the evolution of $\rho_h(z, t)$ in the switched harmonic potential is shown in fig.\ \ref{fig:harmonic_trap_hybrid}.
We observe that the initial state is captured via the hybrid method much more accurately than by the individual DFT-routes.
This trend transfers to the relaxation dynamics, where arguably better results can be achieved than with standard and force-DDFT alone.
Still, hybrid DDFT remains adiabatic, such that effects beyond the adiabatic assumption are not incorporated by construction.
In the considered case, however, this approximation turns out to be reasonable, and the resulting density evolution calculated within DDFT can hence be improved by the combination procedure \eqref{eq:rho_hybrid} as we had shown before for equilibrium DFT.

\section{Conclusions and outlook}
\label{sec:conclusion}

In this work, the recent force-DFT method developed by \citeauthor{Tschopp2022} \cite{Tschopp2022} was compared in-depth to standard DFT and simulation results.
For this, we have reexamined the results of Ref.\ \onlinecite{Tschopp2022} for a hard sphere fluid both in equilibrium at a hard wall as well as for its relaxation dynamics in a switched harmonic trap.
Numerically exact many-body simulations have been carried out to enable the comparison of density profiles from standard and force-DFT calculations with reference data.

We first turned to the prototypical case of subjecting the hard sphere fluid to a hard wall, thereby inducing large density modulations.
As shown by \citeauthor{Tschopp2022} \cite{Tschopp2022} standard and force-DFT are connected via the hard wall contact theorems \eqref{eq:contact_theorem_c} and \eqref{eq:contact_theorem_v} to the compressibility and virial expression of the pressure, respectively, which was exemplified in their work with the Rosenfeld FMT functional and the corresponding Percus-Yevick equation of state.
Here, we have augmented this investigation with numerically accurate density profiles from GCMC simulations, which have been adjusted to replicate the same bulk density as used in both DFT methods.
As expected from the theoretical results of Ref.\ \onlinecite{Tschopp2022}, the numerical contact density is enclosed by the results from standard and from force-DFT and fits more accurately to the former.
More importantly, however, with the GCMC data being available, the comparison could be carried out in this work for the complete inhomogeneous structure of the density profile.
For intermediate distances from the wall, the numerical density profile shows discrepancies to the results of both DFT routes.
In large parts of the system, the GCMC density profile is bracketed by standard and force-DFT results.
In the vicinity of the first density maximum, which is difficult to reproduce in standard DFT \cite{Davidchack2016}, force-DFT yields no systematic improvement.

We next considered the dynamical relaxation of the hard sphere fluid in a harmonic potential when its strength is instantaneously decreased.
In order to complement the force-DDFT results of Ref.\ \onlinecite{Tschopp2022} with numerical data, we have employed EDBD as an accurate dynamical simulation method for hard sphere fluids under nonequilibrium conditions.
Hence, we have initialized $10^4$ EDBD runs with particle configurations obtained via canonical MC simulations and have reproduced the relaxation dynamics after the switching of the harmonic trap.
The total number of particles as given by the integrated density profile has been matched to the DDFT calculations.
We observed that the inaccuracies of the Rosenfeld FMT functional transfer to the dynamical case, such that the numerical density profile lies in between the results of both DDFT routes.
At the center of the trap, force-DDFT overestimates the local value of the density while standard DDFT yields values that are slightly too low.
As the dynamical description with force-DDFT is still adiabatic by construction, the relative relaxation dynamics differs only marginally to that in standard DDFT.

With the previous observations for both routes in equilibrium and in the dynamical case, we have investigated a hybrid method via an appropriate linear interpolation of standard and force-results as was suggested in Ref.\ \onlinecite{Tschopp2022}.
For the hard sphere fluid modeled with the Rosenfeld FMT functional, an interpolation parameter could be found by considering the associated Percus-Yevick results \eqref{eq:eos_py_c} and \eqref{eq:eos_py_v} and their well-known combination \eqref{eq:eos_cs_from_py_combination} to yield the improved Carnahan-Starling equation of state.
We have shown that the application of an analog combination procedure to standard and force-results yields substantially improved density profiles both in equilibrium and in the dynamical scenario.
In equilibrium at the hard wall, we have compared the hybrid method with the Rosenfeld functional both to GCMC data and to density profiles calculated with standard DFT when using the highly accurate White Bear and White Bear MkII functionals.
It was shown that the hybrid Rosenfeld scheme mitigates many deficiencies of the individual DFT routes.
Its deviations from the GCMC data are comparable to those of the standard White Bear and White Bear MkII DFT treatments.

In the time-dependent problem, the hybrid implementation of DDFT captures the relaxation of the hard sphere fluid much better than standard and force-DDFT alone, which we attribute to the more accurate reproduction of the equation of state.
Still, the hybrid scheme is purely adiabatic by construction.
This is an acceptable approximation in the presented case, but will be inappropriate in other dynamical systems.

In the future, it would be interesting to use more accurate functionals such as White Bear and White Bear MkII in force-DFT and in the hybrid method.
As hybrid Rosenfeld DFT already significantly improves upon the individual DFT routes, it is conceivable that a hybrid White Bear (MkII) DFT will lead to a further systematic gain in the accuracy of the resulting density profiles.
Moreover, the method could be useful in other systems that may consist of different particle types than the hard sphere fluid, where the derivation of accurate Helmholtz excess free energy functionals poses an even more difficult problem.
On the other hand, both standard DFT and force-DFT are equivalent if one can start with the exact free energy functional.
Hence carrying out explicitly an investigation for the one-dimensional hard core (``hard rod'') system using Percus' exact functional \cite{Percus1976} as a practical verification of the formal equivalence of both DFT routes could be a worthwhile future research task.
This could be augmented by a force-DFT investigation of the two-dimensional hard disk system, where both highly accurate FMT functionals \cite{Roth2012} as well as highly reliable simulation results \cite{Li2022} have been reported.

From a conceptual point of view, force-DFT opens up the possibility to gain further insight into the inner workings of DFT, especially by making the two-body density correlation function directly accessible.
This could be used, e.g., in an investigation of the hard sphere pair correlations at the contact shell.
Furthermore, one could obtain one-body fluctuation profiles \cite{Eckert2020} such as the local compressibility \cite{Evans2015,Coe2022} from integrating over the two-body pair correlation function.
This offers an alternative way to access this information besides the common parametric differentiation of the density profile.
Of course, standard DFT also allows to compute the pair structure via the inhomogeneous OZ equation, see e.g.\ the work carried out by Dietrich and coworkers \cite{Goetzelmann1996,Mecke1999,Klimpel1999}.
We further point out that higher densities than showcased in this work could be investigated, which becomes a conceptually demanding test case when approaching the freezing transition.
Additionally, more advanced hybrid schemes are conceivable, e.g.\ by using a local mixing parameter $\alpha(\vec{r})$, and from a theoretical perspective, self-consistency of standard and force-DFT could be a useful prerequisite in the derivation of accurate excess free energy functionals.
This is especially interesting from the viewpoint of FMT, where the construction and choice of appropriate nonlocal measures is an ongoing research task \cite{Tarazona2000,Tarazona2002,Tarazona2008}.
One could hope that force- and hybrid DFT shed light on the clearly noticeable deficiencies of FMT and provide aid in the derivation of improved hard sphere functionals.

When dynamics are considered, the prospects arising from the force route are even more promising than in equilibrium.
A fundamental advantage of the force-DDFT formalism is the possibility to include higher orders in the many-body hierarchy.
Recently, \citeauthor{Tschopp2022a} \cite{Tschopp2022a} exploited this idea by considering the dynamics of the two-body density explicitly via its continuity equation.
Applying the adiabatic approximation only at this higher order then yields a systematic extension of standard DDFT that is no longer adiabatic on the one-body level.
Further possibilities to break free of the inherent restrictions of standard DDFT are discussed in Ref.\ \onlinecite{Heras2023}.

\begin{acknowledgments}
  We thank M.\ Coe for providing a DFT library with implementations of the White Bear and White Bear MkII functionals \footnote{The DFT library can be found at \url{https://github.com/marykcoe/cDFT_Package}.}.
  J.\ M.\ Brader and S.\ M.\ Tschopp are thanked for sending us the theoretical data of Ref.\ \onlinecite{Tschopp2022}.
  L.\ L.\ Treffenstädt is acknowledged for sharing his EDBD code and D.\ de las Heras and R.\ Evans for useful discussions.
  This work is supported by the German Research Foundation (DFG) via project number 436306241.
\end{acknowledgments}

\bibliography{bibliography.bib}

\appendix

\section{Chemical potential from the Percus-Yevick equation of state}
\label{appendix:mu}

We briefly give some classical results and point out Ref.\ \onlinecite{Santos2020} for an extensive and well-accessible collection of analytical relations for the hard sphere fluid.
The Percus-Yevick equation of state
\begin{equation}
  \label{eq:eos_py_general}
  P_{c, v} = \frac{\rho_b}{\beta} f_{c, v}(\eta)
\end{equation}
can be obtained either via the compressibility (subscript $c$) or the virial (subscript $v$) route.
The explicit forms of the functions $f_{c, v}(\eta)$ are given in eqs.\ \eqref{eq:eos_py_c} and \eqref{eq:eos_py_v} in the main text.

We consider the Helmholtz free energy $F$ and insert eq.\ \eqref{eq:eos_py_general}, which yields
\begin{equation}
  F = - \int \diff V P = \frac{N}{\beta} \int \diff \rho_b \frac{f(\eta(\rho_b))}{\rho_b}.
\end{equation}
The chemical potential is then obtained via
\begin{equation}
  \begin{split}
    \mu &= \frac{\partial F}{\partial N} = \frac{\partial F/V}{\partial \rho_b}\\
    &= \frac{1}{\beta} \left( f(\eta) + \int \diff \rho_b \frac{f(\eta(\rho_b))}{\rho_b} \right).
  \end{split}
\end{equation}
Thus,
\begin{align}
  \label{eq:mu_c}
  \beta \mu_c &= \ln(\rho_b) + f_c(\eta) + \frac{3}{2 (1 - \eta)^2} - \ln(1 - \eta) - \frac{5}{2},\\
  \label{eq:mu_v}
  \beta \mu_v &= \ln(\rho_b) + f_v(\eta) + \frac{6 \eta}{1 - \eta} + 2 \ln(1 - \eta) - 1.
\end{align}

\end{document}